# Electric self inductance of quasi-2D magnetic-dipolar-mode ferrite disks

M. Sigalov, E.O. Kamenetskii, and R. Shavit

Ben-Gurion University of the Negev, Beer Sheva 84105, Israel

## Abstract

An electric current flowing around a loop produces a magnetic field and hence a magnetic flux through the loop. The ratio of the magnetic flux to the electric current is called the (magnetic) self inductance. Can there be a dual situation with a magnetic current flowing around a loop and producing an electric field and hence an electric flux through the loop? Following the classical electrodynamics laws an answer to this question should be negative. Nevertheless special spectral properties of magnetic dipolar modes in a quasi-2D ferrite disk show there are the double-valued-function loop magnetic currents which may produce eigen electric fields and hence eigen electric fluxes through the loop. In this case one can definitely introduce the notion of the electric self inductance as the ratio of the electric flux to the magnetic current. In this paper we show experimentally that in the magnetic-dipolar-mode ferrite disks there exist eigen electric fluxes. These fluxes are very sensitive to permittivity parameters of materials abutting to the ferrite disk. Dielectric samples above a ferrite disk with a higher permittivity than air confine the electric field closely outside the ferrite, thereby changing the loop magnetic currents and thus transforming the magnetic-dipolar-mode oscillating spectrum.

## 1. Introduction

Magnetostatic (MS) ferromagnetism has a character essentially different from exchange ferromagnetism [1, 2]. This statement finds strong confirmation in confinement phenomena of magnetic-dipolar-mode (MDM) oscillations. The dipole interaction provides us with a long-range mechanism of interaction, where a magnetic medium is considered as a continuum. In a general case, both short-range exchange and long-range dipole-dipole (magnetostatic) interactions contribute to eigenfrequencies of the collective spin excitation. The importance of MS energy increases gradually as the particle size increases. MDMs are not eigen modes of magnetization. Contrary to an exchange spin wave, in magnetic-dipolar waves the local fluctuation of magnetization does not propagate due to interaction between the neighboring spins. When field differences across the sample become comparable to the bulk demagnetizing fields the local-oscillator approximation is no longer valid and there should be certain propagating fields – the MS fields – which cause and govern propagation of magnetization fluctuations. In other words, space-time magnetization fluctuations are corollary of the propagating MS fields, but there are no magnetization waves. The boundary conditions should be imposed on the MS field and not on the RF magnetization.

Ferrite MDM samples are well localized in space, their extension is assumed to be much smaller than the variation length of the electromagnetic field. It is well known that in a general case of small (compared to the free-space electromagnetic-wave wavelength) samples made of media with strong temporal dispersion, the role of displacement currents in Maxwell equations can be negligibly small, so oscillating fields are the quasistationary fields. In small samples with strong temporal dispersion of the permeability tensor: $\vec{\mu} = \vec{\mu}\,(\omega)$, variation of the electric energy is negligibly small compared to variation of the magnetic energy and so one can neglect the electric displacement current in Maxwell equations [3]. These magnetic samples can exhibit the MS resonance behavior in microwaves [4 – 7]. For such resonance MDMs, the quasimagnetostatic fields are described by a scalar magnetostatic potential $\psi$. This potential does not have the same physical meaning as in a situation of pure magnetostatics. In fact, there is the MS-potential *wave* function $\psi(\vec{r},t)$ describing the resonant behavior in a small magnetic object.

Can the MS-potential wave function be used for comprehensive description of MDM oscillations in small magnetic samples? In attempts to use the notion of scalar wave function $\psi(\vec{r},t)$ for the spectral analysis one becomes faced with evident contradictions with the dynamical Maxwell equations. This fact can be perceived, in particular, from the remarks made by McDonalds [8]. For MS resonances in small magnetic objects one neglects the electric displacement current: $\frac{\partial \vec{D}}{\partial t} = 0$. From Maxwell equation (the Faraday law), $\nabla \times \vec{E} = -\frac{1}{c}\frac{\partial \vec{B}}{\partial t}$, one obtains $\nabla \times \frac{\partial \vec{E}}{\partial t} = -\frac{1}{c}\frac{\partial^2 \vec{B}}{\partial t^2}$. If a sample does not posses any dielectric anisotropy, we have $\frac{\partial^2 \vec{B}}{\partial t^2} = 0$. It follows that the magnetic field in small resonant magnetic objects vary linearly with time. This leads, however, to arbitrary large fields at early and late times, and is excluded on physical grounds. An evident conclusion suggests itself at once: the magnetic (for MS resonances) fields are constant quantities. This contradicts to the fact of temporally dispersive media and any resonant conditions. Another conclusion is more unexpected: for a case of MS resonances the Faraday law is not valid. Concerning the MS-wave propagation effects, it was disputed also in [9 – 11] that from a classical electrodynamics point of view one does not have a physical mechanism describing the effect of transformation of the curl electric field to the potential magnetic field. Also the gauge transformation in this derivation does not fall under the known gauge transformations, neither the Lorentz gauge nor the Coulomb gauge, and cannot formally lead to the wave equation.

MS waves can propagate only due to ferrite-medium confinement phenomena [12, 13]. So the power flow density of MS waves should be considered via an analysis of the mode spectral problem in a certain waveguide structure, but not based on an analysis of the wave propagation in a boundless magnetic medium. It means that the MS-wave power flow density should have the only physical meaning as a norm of a certain propagating mode in a magnetic waveguide structure. As a necessary consequence, this leads to the question of orthogonality and completeness of MDMs in such a waveguide. In this problem MS-potential function $\psi$ acquires a special physical meaning as a scalar wave function in a Hilbert functional space. In an analysis of the MDM oscillating spectra, a ferrite-disk particle is considered as a section of an axially magnetized ferrite rod. For a flat ferrite disk, having a diameter much bigger than a disk thickness, one can successfully use separation of variables for the MS-potential wave function [14, 15]. A similar way of separation variables is used, for example, in solving the electromagnetic-wave spectral problem in dielectric disks [16].

It was shown [10, 11, 14, 15] that for MDMs in a ferrite disk one has evident quantum-like attributes. The spectrum is characterized by *energy eigenstate* oscillations. The energy eigenvalue problem is defined by the differential equation:

$$\hat{F}_{\perp}\tilde{\varphi}_p = E_p\tilde{\varphi}_p\,, \tag{1}$$

where $\tilde{\varphi}_p$ is a dimensionless membrane MS-potential wave function and $E_p$ is the normalized average (on the RF period) density of accumulated magnetic energy of mode *p*. A two-dimensional ("in-plane") differential operator $\hat{F}_{\perp}$ in Eq. (1) is defined as:

$$\hat{F}_{\perp} = \frac{g_p}{16\pi}\mu\nabla_{\perp}^2\,, \tag{2}$$

where $\nabla_{\perp}^2$ is the two-dimensional (with respect to cross-sectional coordinates) Laplace operator, $\mu$ is a diagonal component of the permeability tensor, and $g_p$ is a dimensional normalization coefficient for mode *p*. Operator $\hat{F}_{\perp}$ is a positive definite differential operator for negative quantity $\mu$. The energy density of mode *p* is determined as

$$E_p = \frac{g_p}{16\pi}\left(\beta_{z_p}\right)^2, \tag{3}$$

where $\beta_{z_p}$ is the propagation constant of mode *p* along disk axis *z*. For MDMs in a ferrite disk at a constant frequency one has the energy orthonormality condition:

$$(E_p - E_{p'})\int_S \tilde{\varphi}_p\tilde{\varphi}_{p'}^* dS = 0\,, \tag{4}$$

where *S* is a cylindrical cross section of an open disk. For constant frequency, different mode energies one has at different quantities of a bias magnetic field. From the principle of superposition of states, it follows that wave functions $\tilde{\varphi}_p$ ( $p = 1,2,...$), describing our "quantum" system, are "vectors" in an abstract space of an infinite number of dimensions – the Hilbert space. In quantum mechanics, this is the case of so-called energetic representation, when the system energy runs through a discrete sequence of values [17, 18].

It was shown, however, that because of the boundary condition on a lateral surface of a ferrite disk, membrane functions $\tilde{\varphi}_p$ cannot be considered as single-valued functions [10, 11]. This fact raises a question about validity of the energy orthogonality relation for the MDMs. The most basic implication of the existence of a phase factor in $\tilde{\varphi}$ is operative in the case on the border ring region. In order to cancel the "edge anomaly", the boundary excitation must be described by chiral states. Because of such chiral states, the double-valued-function loop magnetic currents occur. These currents may produce eigen electric fields and hence eigen electric fluxes through the loop [10, 11]. In this case one can definitely introduce the notion of the electric self inductance as the ratio of the electric flux to the magnetic current.

In this paper we show experimentally that in a quasi-2D ferrite disk with MDM oscillations there exist eigen electric fluxes. One has an evidence for strong sensitivity of the MDM spectral characteristics to the dielectric-load permittivity quantity. The oscillating MDMs can be characterized by certain electric self-inductance parameters. We start with a brief description of such non trivial notions as eigen electric fluxes, persistent magnetic currents and anapole moments in MDM ferrite disks**.** These notions are partly known from the literature. Nevertheless, an initial classification of unique topological effects in the MDM ferrite disks could be very useful for proper characterization of our experimental results.

## 2. Eigen electric fluxes, persistent magnetic currents and anapole moments in MDM ferrite disks

The topological effects in the MDM ferrite disk are manifested through the generation of relative phases which accumulate on the boundary wave functions $\delta_\pm$ [10, 11]:

$$\delta_\pm \equiv f_\pm e^{-iq_\pm \theta}, \tag{5}$$

where $\theta$ is an azimuth coordinate in a cylindrical coordinate system. The quantities $q_\pm$ are equal to $\pm l \frac{1}{2}$, $l = 1, 3, 5, \ldots$ For amplitudes *f* we have $f_+ = -f_-$ with normalization $\left| f_\pm \right|$ = 1. To preserve the single-valued nature of the membrane functions of the MDM oscillations, functions $\delta_\pm$ must change its sign when a disk angle coordinate $\theta$ is rotated by $2\pi$ so that $e^{-iq_\mp 2\pi} = -1$. A sign of a full chiral rotation, $q_+\theta = \pi$ or $q_-\theta = -\pi$, should be correlated with a sign of the parameter $i\mu_a$ – the off-diagonal component of the permeability tensor $\vec{\mu}$. This becomes evident from the fact that a sign of $i\mu_a$ is related to a precession direction of a magnetic moment $\vec{m}$. In a ferromagnetic resonance, the bias field sets up a preferential precession direction. It means that for a normally magnetized ferrite disk with a given direction of a DC bias magnetic field, there are two types of resonant oscillations, which we conventionally designate as the (+) resonance and the (–) resonance. For the (+) resonance, a direction of an edge chiral rotation coincides with the precession magnetization direction, while for the (–) resonance, a direction of an edge chiral rotation is opposite to the precession magnetization direction.

For a ferrite disk with *r* and $\theta$ in-plane coordinates, the total MS-potential membrane function $\tilde{\varphi}$ is represented as a product of two functions [10, 11]:

$$\tilde{\varphi} = \tilde{\eta}(r,\theta)\, \delta_\pm, \tag{6}$$

where $\tilde{\eta}(r,\theta)$ is a single-valued membrane function, and $\delta_\pm$ is a double-valued edge (spin-coordinate-like) function. We may introduce a "spin variable" $\sigma$, representing the orientation of the "spin moment" and two double-valued wave functions, $\delta_+(\sigma)$ and $\delta_-(\sigma)$. The two wave functions are normalized and mutually orthogonal, so that they satisfy the equations $\int \delta_+^2(\sigma)\, d\sigma = 1$, $\int \delta_-^2(\sigma)\, d\sigma = 1$, and $\int \delta_+(\sigma)\, \delta_-(\sigma)\, d\sigma = 0$. A membrane wave function $\tilde{\varphi}$ is then a function of three coordinates, two positional coordinates such as $r, \theta$, and the "spin coordinate" $\sigma$. For the positional wave function $\tilde{\eta}(r,\theta)$, which is a solution of the Walker equation for a ferrite disk with the so-called essential boundary conditions [15], there could be two equiprobable solutions for the membrane wave functions: $\tilde{\varphi}_+ = \tilde{\eta}(r,\theta)\, \delta_+(\sigma)$ and $\tilde{\varphi}_- = \tilde{\eta}(r,\theta)\, \delta_-(\sigma)$.

The geometrical phase factor is not single-valued under continuation around a circuit and can be correlated with the vector potential. The vector potential is considered to be nonobservable in Maxwellian electromagnetism. At the same time, the vector potential can be observable in the Aharonov-Bohm [19] or Aharonov-Casher [20] effects, but only via its line integral, not pointwise. From an analysis in [10, 11] it follows that restoration of singlevaluedness (and, therefore, Hermicity) of the MDM spectral problem in a ferrite disk is due to a line integral $\oint_C \left(i\vec{\nabla}_\theta \delta_\pm\right)\left(\delta_\pm^\circ\right)^* dC$, where $C = 2\pi\,\Re$ is a contour surrounding a cylindrical ferrite core, $\Re$ is a ferrite disk radius, and $\delta_\pm^\circ$ are the boundary wave functions for the conjugate problem. Such a line integral is an observable quantity which can be represented as an integral of a certain vector potential:

$$i\Re\int_0^{2\pi}[(\vec{\nabla}_\theta\delta_\pm)(\delta_\pm^\circ)^*]_{r=\Re}\,d\theta \equiv \oint_C \left(\vec{A}_\theta^m\right)_\pm \cdot d\vec{C} = 2\pi q_\pm \tag{7}$$

One can see that in the problem under consideration the Berry's phase [21] is generated from the broken dynamical symmetry. To compensate for sign ambiguities and thus to make wave functions single valued we added a vector-potential-type term to the MS-potential Hamiltonian. A circulation of vector $\vec{A}_\theta^m$ should enclose a certain flux. The corresponding flux of pseudo-electric field $\vec{\in}$ (the gauge field) through a circle of radius $\Re$ is obtained as:

$$\int_S \left(\vec{\in}\right)_\pm \cdot d\vec{S} = \oint_C \left(\vec{A}_\theta^m\right)_\pm \cdot d\vec{C} = \left(\Xi^e\right)_\pm, \tag{8}$$

where $\left(\Xi^e\right)_\pm$ is the flux of pseudo-electric field. There should be the positive and negative fluxes. These different-sign fluxes should be inequivalent to avoid the cancellation. Superscript $m$ in the vector-potential term means that there is the physical quantity associated with the magnetization motion. A magnetic moment moving along contour $C$ in the gauge field feels no force and undergoes the Aharonov-Bohm-type interference effect.

For every MDM, the total solution of the MS-potential wave function in a cylindrical ferrite disk is represented as

$$\psi(r,\theta,z) = \xi(z)\tilde{\varphi}(r,\theta), \tag{9}$$

while a single-valued membrane function is represented as

$$\tilde{\eta}(r,\theta) = R(r)\phi(\theta). \tag{10}$$

In these equations $\xi(z)$ describes an axial distribution of the MS-potential wave function, $R(r)$ is described by the Bessel functions and $\phi(\theta) \sim e^{-i\nu\theta}$, $\nu = \pm1, \pm2, \pm3....$

Based on Eqs. (7) – (10) and taking into account an analysis of self-adjointness of the MDM differential operators [10, 11], we can represent the total flux of pseudo-electric field for every MDM in a ferrite disk as

$$\Xi^e_{total} = 2\pi\, q_\pm R^2_{r=\Re} \int_0^d \xi(z)\, dz \ , \tag{11}$$

where *d* is a disk thickness.

In the topological effects of the generation of relative phases which accumulate on the boundary wave function $\delta_\pm$, the quantity $\nabla_\theta \delta_\pm$ can be considered as the velocity of an irrotational "border" flow:

$$\left(\vec{\mathrm{v}}_\theta\right)_\pm \equiv \vec{\nabla}_\theta \delta_\pm \,. \tag{12}$$

In such a sense, functions $\delta_\pm$ are the velocity potentials. Circulation of $\left(\vec{\mathrm{v}}_\theta\right)_\pm$ along a contour $C$ is equal to $\oint_C \left(\vec{\mathrm{v}}_\theta\right)_\pm \cdot d\vec{C} = \Re \int_0^{2\pi} \nabla_\theta \delta_\pm \, d\theta = -2 f_\pm$. Taking into account the "orbital" function $\phi(\theta)$, we may consider the quantity $\left[\vec{\nabla}_\theta \left(\phi\, \delta_\pm\right)\right]_{r=\Re}$ as the total ("orbital" and "spin") velocity of an irrotational "border" flow:

$$\left(\vec{V}_\theta\right)_\pm \equiv \left[\vec{\nabla}_\theta \left(\phi\, \delta_\pm\right)\right]_{r=\Re} . \tag{13}$$

It is evident that

$$\left(\vec{V}_\theta\right)_\pm = -i \frac{\left(\nu + q_\pm\right) f_\pm}{\Re} e^{-i\left(\nu+q_\pm\right)\theta} \vec{e}_\theta \,. \tag{14}$$

For a given membrane function $\tilde{\eta}$ and given *z*-distribution of the MS potential, $\xi(z)$, we can define now the strength of a vortex of a whole disk as

$$s^e_\pm \equiv R_{r=\Re} \int_0^d \xi(z) dz \oint_C \left(\vec{V}_\theta\right)_\pm \cdot d\vec{C} = \Re R_{r=\Re} \int_0^d \xi(z) dz \int_0^{2\pi} \left(\vec{V}_\theta\right)_\pm \cdot \vec{e}_\theta \, d\theta = -2 f_\pm R_{r=\Re} \int_0^d \xi(z) dz \,. \tag{15}$$

The quantity $\left(\vec{V}_\theta\right)_\pm$ has a clear physical meaning. In the spectral problem for MDM ferrite disks, non-singlevaluedness of the MS-potential wave function appears due to a border term which arises from the demand of conservation of the magnetic flux density on a lateral surface of a disk [10, 11]. This border term is defined as $-i\mu_a \left(H_\theta\right)_{r=\Re}$, where $\mu_a$ is the off-diagonal component of the permeability tensor and $\left(H_\theta\right)_{r=\Re}$ is an annual magnetic field on the border circle. It is evident that

$$\left(\left(\vec{H}_\theta(z)\right)_\pm\right)_{r=\Re} = -\xi(z) \left(\vec{V}_\theta\right)_\pm . \tag{16}$$

We define now an angular moment $\vec{a}^e_\pm$:

$$a_{\pm}^{e} \equiv \int_{0}^{d} \oint_{C} \left[ -i\mu_a \left( H_\theta \right)_{r=\Re} \right] \vec{e}_\theta \cdot d\vec{C}\, dz = i\mu_a\, s_{\pm}^{e}. \tag{17}$$

This angular moment can be formally represented as a result of a circulation of a quantity, which we call a density of an effective boundary magnetic current $\vec{i}^{\,m}$:

$$a_{\pm}^{e} = 4\pi \int_{0}^{d} \xi(z)\, dz \oint_{C} \vec{i}_{\pm}^{\,m} \cdot d\vec{C} \ , \tag{18}$$

where

$$\vec{i}_{\pm}^{\,m} \equiv \rho^{m} \left( \vec{V}_\theta \right)_{\pm} \tag{19}$$

and

$$\rho^{m} \equiv i \frac{\mu_a}{4\pi} R_{r=\Re}\,. \tag{20}$$

In our continuous-medium model, a character of the magnetization motion becomes apparent via the gyration parameter $\mu_a$ in the boundary term for the spectral problem. There is magnetization motion through a non-simply-connected region. On the edge region, the chiral symmetry of the magnetization precession is broken to form a flux-closure structure. The edge magnetic currents can be observable only via its circulation integrals, not pointwise. This results in the moment oriented along a disk normal. It was shown experimentally [22] that such a moment has a response in an external RF electric field. This clarifies a physical meaning of a superscript "e" in designations of $s_{\pm}^{e}$ and $a_{\pm}^{e}$. In a ferrite disk particle, the vector $\vec{a}^{e}$ is an electric moment characterizing by special symmetry properties.

An electric moment $a_{\pm}^{e}$ is characterized by the anapole-moment properties. This is a certain-type toroidal moment. Some important notes should be given here to characterize properties of moment $\vec{a}^{e}$. From classical consideration it follows that for a given electric current $\vec{i}^{\,e}$, a magnetic dipole moment is described as $\vec{m} = \frac{1}{2c}\int \vec{r} \times \vec{i}^{\,e}\, dv$, while the toroidal dipole moment is described as $\vec{t} = \frac{1}{3c}\int \vec{r} \times \left( \vec{r} \times \vec{i}^{\,e} \right) dv$ (see e.g. [23]). When we introduce the notion of an elementary magnet: $\vec{M} \equiv \vec{r} \times \vec{i}^{\,e}$, we can represent the toroidal dipole moment as a linear integral around a loop: $\vec{t} = \frac{1}{3c}\int \vec{r} \times \vec{M}\, dl$. It is considered as a ring of elementary magnets $\vec{M}$. In this formulation, it is clear that a toroidal moment is parity odd and time reversal odd. In a case when $\vec{M}$ is time varying, one has a magnetic current $\vec{i}^{\,m} \equiv \frac{\partial \vec{M}}{\partial t}$ and a linear integral of this current around a loop defines a moment which is parity odd and time reversal even. This is the case of an anapole moment $\vec{a}^{e}$ which has the symmetry of an electric dipole. From classical point of view such a definition presumes no azimuth variations of loop magnetic current $i^{m}$. In our case, however, for oscillating MDMs one has the azimuth varying ring magnetic current.

The magnetic current $i^m$ is described by the double valued functions. This results in appearance of an anapole moment $\vec{a}^e$.

## 3. Characterization of interaction of MDMs with the external electromagnetic fields

Microwave experiments [4, 6, 7] give evidence for the multiresonance MDM oscillations in ferrite disks. The absorption peak positions corresponding to these MDMs depend on the ferrite material parameters and the disk geometry but not on the type of a cavity. This statement was confirmed more in details in experimental paper [22].

The mechanism of excitation of the multiresonance MDM spectra by the cavity electromagnetic fields is, however, not so obvious. Following the theory in [9 – 11] one sees that the main factor of interaction of MDMs with the cavity electromagnetic fields is caused by the presence of topological singularities on the ferrite disk surfaces. These topological singularities are described by the double valued functions and the main aspects concern the question how one can describe effective resonance interactions between the double-valued-function edge states and single-valued-function cavity electromagnetic fields. The mechanism of such an interaction we illustrate here by the following qualitative models.

Suppose that we have the (+) resonance which is characterized by azimuth number $q = +\frac{1}{2}$. Fig. 1 (a) shows the double-valued functions $\delta_+(\theta')$ and $\nabla_{\theta'}\delta_+(\theta')$. Here we use designation $\theta'$ to distinguish the "spin" angular coordinate from a regular angle coordinate $\theta$. Because of the edge-function chiral rotation [10, 11], for the (+) resonance one has to select only positive derivatives: $\frac{\partial \delta_+}{\partial \theta'} > 0$. The corresponding parts of the graphs are distinguished in Fig. 1 (a) by bold lines. Figs. 1 (b) and (c) give two cases of the single-valued-function cavity field $F(\theta)$ which may lead to resonance interactions with the double-valued-function edge state. It is evident that in a case of Fig. 1 (b), a positive half of function $\nabla_{\theta'}\delta_+(\theta')$ is phased for a resonance interaction with the positive halves of function $F(\theta)$, while in a case of Fig. 1 (c) a positive half of function $\nabla_{\theta'}\delta_+(\theta')$ is phased for a resonance interaction with the negative halves of function $F(\theta)$. Interaction with the cavity field shown in Fig. 1 (b) can be characterized as the "resonance absorption" while interaction with the cavity field shown in Fig. 1 (c) – as the "resonance repulsion". Both types of interactions are equiprobable and can be exhibited separately. One may also expect that in a certain situation transitions between these two resonance behaviors can be demonstrated. A similar model can be used for illustration of a possible mechanism of interactions in a case of the (–) resonance. For $q = -\frac{1}{2}$, the double-valued functions $\delta_-(\theta')$ and $\nabla_{\theta'}\delta_-(\theta')$ are shown in Fig. 2 (a). Here we have to select only negative derivatives: $\frac{\partial \delta_-}{\partial \theta'} < 0$. In this case, interaction with the cavity field shown in Fig. 2 (b) can be characterized as the "resonance absorption" and interaction with the cavity field shown in Fig. 2 (c) – as the "resonance repulsion".

An interaction of the anapole moment of a ferrite disk with the cavity RF electric field was clearly demonstrated in [22]. Following the model shown in Figs. 1 and 2 one can describe this interaction by the average quantity $\frac{1}{2\pi}\int_0^{2\pi}\left(\vec{a}^e \cdot \vec{E}\right) d\theta$, where $\vec{E}$ is the cavity electric field. The positive quantity of this integral corresponds to the "resonance absorption" while for a negative integral one has the "resonance repulsion". These two behaviors we demonstrate in the following experiments.

In experiments, we used a disk sample of a diameter $2\Re = 3\,\text{mm}$ made of the YIG film on the GGG substrate (the YIG film thickness $d = 49.6\,\text{mkm}$, saturation magnetization $4\pi M_0 = 1880\,\text{G}$, linewidth $\Delta H = 0.8\,\text{Oe}$; the GGG substrate thickness is 0.5 mm). A normally magnetized ferrite-disk sample was placed in a rectangular waveguide cavity with the $TE_{102}$ resonant mode. The disk axis was oriented along the waveguide *E*-field (see Fig. 3). We analyzed the MDM spectra for the DC magnetic field variation at certain constant frequencies. To investigate behaviors of the "resonance repulsion" and "resonance absorption" and transition between these behaviors we analyzed the ferrite disk spectra measured at different frequencies. The multiresonance spectral pictures are shown in Fig. 4 (a). These frequencies, $f_1$, $f_{2,}$ and $f_3$, correspond to different positions on the resonance curve of the cavity [see Fig. 4 (b)]. The spectral peaks of the MDM oscillations one obtains at certain permeability tensor parameters [11, 15]. Since the permeability tensor parameters are dependent both on frequency and a bias magnetic field, we are able to match the peak position by small variations of a bias field. The bias magnetic fields corresponding to the first peaks are adduced in Fig. 4 (a). The digits characterize the MDM numbers.

In Fig. 4 (a), the spectrum corresponding to $f_1$ represents the "resonance repulsion" behavior. At the same time, the spectrum corresponding to $f_3$, clearly demonstrates the "resonance absorption" behavior. It becomes evident that the spectrum corresponding to $f_2$, shows the transitions between the "resonance repulsion" and "resonance absorption". A qualitative explanation of the observed three cases could be the following. Since at frequency $f_1$ the cavity is "viewed" by the incoming signal as an active load, one can clearly observe the "resonance repulsion" due to a ferrite disk. Contrary, at frequency $f_3$ the cavity is characterized mainly as a reactive load. In this case one observes the "resonance absorption" behavior. At frequency $f_2$ both cases are mixed and a transitional behavior takes place.

The main conclusion following from the above consideration is that there exists a real mechanism of the effective resonance interactions between the double-valued-function edge states and single-valued-function electromagnetic fields. The character of the MDM resonance responses may depend on the EM field structure but the resonance peak positions are fixed for a given ferrite disk. These results will allow us to analyze the electric self-inductance properties of MDM ferrite disks.

## 4. Experimental evidence for the electric self-inductance properties of MDM ferrite disks

In the above experiments we analyzed the MDM excitation by the cavity fields as a result of interaction of the eigen electric (anapole) moments with the external RF electric fields. An anapole moment is created by a loop magnetic current. It can be supposed that an equivalent alternative mechanism of excitation which describes an induction of the loop magnetic current by the time varying electric-field flux can also be used. A dual effect of induction of an electric displacement current in a dielectric-disk resonator due to the Faraday law was analyzed in [24]. The physics of this effect, however, is completely different from our case.

First of all, let us define the notion of the electric self inductance for MDMs. For every MDM in a ferrite disk, we define the electric self inductance as the ratio of the electric flux to the loop magnetic current. Based on Eq. (11) we express the eigen electric flux for a given mode *p* as

$$\left(\Xi^e\right)_p = 2\pi\, q_\pm \left(R^2_{r=\Re}\right)_p \int_0^d \xi_p\, dz\ . \qquad (21)$$

Following Fig. 1 or Fig. 2 and based on Eqs. (19) and (20), the time average magnetic current for mode *p* is expressed as

$$\left(i^m_{average}\right)_p = i\frac{(\mu_a)_p}{4\pi}\left(R_{r=\Re}\right)_p \frac{1}{\pi}\int_0^{\pi}\nabla_{\theta'}\delta\, d\theta' = i\frac{(\mu_a)_p}{2\pi^2}\left(R_{r=\Re}\right)_p . \qquad (22)$$

We define the electric self inductance for a given mode *p* as

$$\left(L^e\right)_p = \frac{\left(\Xi^e\right)_p}{\left(i^m_{average}\right)_p} = \frac{4\pi^3}{(\mu_a)_p}\left(R_{r=\Re}\right)_p q_{\pm}\int_0^d \xi_p\, dz . \qquad (23)$$

One may presuppose that this is a constant quantity for every MDM in a ferrite disk.

For experimental evidence of the electric self-inductance properties of MDM ferrite disks we suggest the following measurements. Suppose that we put a dielectric sample above a ferrite disk. Because of the electric flux $\Xi^e$ originated from a ferrite disk, the electric charges will be induced on surfaces of a dielectric sample. These surface charges create their own electric flux which will pierce a ferrite disk in an opposite direction with respect to the electric flux $\Xi^e$. As a result of the reduction of the total electric flux, the loop magnetic current in a ferrite disk will be reduced as well, preserving a constant quantity of $L^e$ for a given mode *p*:

$$\left(L^e\right)_p = \frac{\left(\Xi^e\right)_p'}{\left(i^m_{average}\right)_p'} . \qquad (24)$$

Here $\left(\Xi^e\right)_p'$ and $\left(i^m_{average}\right)_p'$ are, respectively, the total electric flux piercing a ferrite disk and the loop magnetic current corresponding to mode *p* when a dielectric sample is put on. For a given mode structure, the requirement of constancy of quantity $L^e$ will lead to a shift of the MDM spectra to lower quantities of a modulus of $\mu_a$ [see Eq. (22)]. In experiments, this should give an evidence for shift of the MDM position in the spectrum to a lower bias field (at a constant frequency) or to a higher frequency (at a constant bias field). Obviously, this shift will be dependent on the permittivity parameter of a dielectric sample. For larger permittivity parameter $\varepsilon_r$ of a dielectric sample, the shift should be larger as well. Moreover, since for a given geometry of a dielectric sample, the flux $\left(\Xi^e\right)_p'$ will abate linearly as $\varepsilon_r$ increases and the current $\left(i^m_{average}\right)_p'$ decreases linearly as a modulus of $\mu_a$ decreases, there should be a linear dependence between $\varepsilon_r$ and a modulus of $\mu_a$ for preserving constant quantity $\left(L^e\right)_p$. In our experiments, we clearly observe this effect for enough small quantities $\varepsilon_r$.

Contrary to the cavity structure shown in Fig. 3, in these experiments we will use a short-wall rectangular waveguide without an entering iris. In this case we can analyze the MDM spectra with respect to the frequency at constant DC magnetic field. We work at a bias magnetic field $H_0 = 4900$ Oe. We start with an analysis of the spectra of a ferrite disk without any dielectric loading at different disk positions in a waveguide. The forms of the MDM resonance peaks may be dependent on the EM field structure and, certainly, the effectiveness of excitation of the modes (the MDM amplitudes) can be different for different disk positions in a waveguide. At the same time, as we discussed above, the resonance peak positions should be fixed for a given ferrite disk. Fig. 5 (a) shows the MDM spectra for different disk positions in a waveguide. These positions are clarified in Fig. 5 (b). In Figs. 6 (a) and 6 (b) one can see more detailed spectral pictures at the positions numbered 1, 6 and 3, 4, respectively. A shift of the disk may slightly perturb the cavity field structure. This perturbation, however, does not influence on the resonance peak positions. In fact, one can see that the peak positions in the MDM spectrum does not depend on the EM field structure in a waveguide.

A ferrite disk is placed on a GGG substrate which has the dielectric permittivity parameter of $\varepsilon_r = 15$. Now we put dielectric samples above a ferrite disk (see Fig. 7). There are dielectric disks of a diameter 3 mm and thickness 2 mm. We used a set of disks of commercial microwave dielectric (non magnetic) materials with the dielectric permittivity parameters of $\varepsilon_r = 3.4$ (RO4003; Rogers Corporation), $\varepsilon_r = 6.15$ (RO3006; Rogers Corporation), $\varepsilon_r = 15$ (K-15; TCI Ceramics Inc), $\varepsilon_r = 30$ (K-30; TCI Ceramics Inc), $\varepsilon_r = 50$ (K-50; TCI Ceramics Inc), and $\varepsilon_r = 100$ (K-100; TCI Ceramics Inc). It can be supposed that an insertion of dielectric samples changes the cavity field structure but, following the results shown in Fig. 5, this field variation should not influence on the absorption peak position. We, however, observed strong variations of the spectral pictures when dielectric disks were placed above a ferrite disk. These transformations of the spectra become most evident when we match (by proper small shifts of the bias magnetic fields) positions of the first peaks in the spectra. From Fig. 8 one can see that as the dielectric permittivity parameter of a dielectric sample increases, the frequency shift of the mode peak position increases as well.

A more detailed picture is shown in Figs. 9, 10 and 11 for the modes 3, 5, and 7, respectively. It is evident that for a given mode (with the mode number more than one) the peak shifts are small compared to the frequency distances between the peak of this mode and the first mode in the oscillating spectra. In a presupposition that the peak shifts are very small one can use the perturbation method assuming that the mode spectral portrait does not change under the perturbation. In this case we can use Eq. (23) for the self inductance characterization.

To analyze the electric self-inductance properties we have to investigate relations between parameters $\varepsilon_r$ and $\mu_a$ for given modes. Fig. 12 shows dependences between $\varepsilon_r$ and a modulus of $\mu_a$ for modes 3, 5 and 7. The quantities $\mu_a$ we calculated for given parameters of a ferrite material, a given bias magnetic field, and for frequencies corresponding to the mode peak positions in Figs. 9, 10, and 11. It is worth noting that for enough small quantities $\varepsilon_r$ of a dielectric sample (for the permittivity less than 30 in our studies) we have almost linear character of dependence between $\varepsilon_r$ and $\mu_a$. For such small permittivity parameters, the Eq. (24) can be applicable.

## 5. Conclusion

Our results show that external RF electric fields can induce loop magnetic currents in MDM ferrite disks. To a certain extent, one can consider this as a dual case with respect to an induction of loop electric currents in dielectric-disk resonators. The physics of these two effects, however, is completely different. In this paper we analyzed the mechanism of interaction of MDMs in a quasi-2D ferrite disk with the external RF fields. This is a special mechanism of interaction between the double-valued and single-valued functions. The double-valued functions describe the MDM geometrical phase factor on the lateral boundary of a ferrite disk, while the single-valued functions describe the MDM membrane oscillations. We showed that due to this mechanism one has an evidence for the electric self-inductance properties of MDM ferrite disks.

It becomes evident that in the MDM oscillating problem, the permittivity parameters of a dielectric layer abutting to the ferrite are the material parameters that influence on the spectral characteristics of MDM oscillations. At the same time, an additional dielectric loading is a small perturbation parameter for the original spectral characteristics.

**Figure captions**

Fig. 1. Interaction of the double-valued-function edge state with the single-valued-function cavity electromagnetic fields in a case of the (+) resonance. (a) Double-valued edge function; (b) the cavity field at the "resonance absorption" interaction; (c) the cavity field at the "resonance repulsion" interaction.

Fig. 2. Interaction of the double-valued-function edge state with the single-valued-function cavity electromagnetic fields in a case of the (–) resonance. (a) Double-valued edge function; (b) the cavity field at the "resonance absorption" interaction; (c) the cavity field at the "resonance repulsion" interaction.

Fig. 3. Rectangular cavity with an inserted ferrite disk

Fig. 4. Experimental evidence for the "resonance repulsion", "resonance absorption", and transitional behaviors. (a) The multiresonance spectral pictures for the "resonance repulsion" (at frequency $f_1$), "resonance absorption" (at frequency $f_3$), and transitional behavior (at frequency $f_2$); (b) frequencies $f_1$, $f_{2,}$ and $f_3$ on the resonance curve of the cavity.

Fig. 5. The spectra of a ferrite disk without dielectric loading. (a) The MDM spectra for different disk positions; (b) disk positions in a waveguide.

Fig. 6. The spectra of a ferrite disk without dielectric loading (detailed spectral pictures). (a) The MDM spectra for positions numbered 1 and 6; (b) the MDM spectra for positions numbered 3 and 4.

Fig. 7. A ferrite disk on a GGG substrate with a dielectric sample.

Fig. 8. The frequency shift of the mode peak positions for a ferrite disk with dielectric loading. The first peaks in the spectra are matched by proper correlations of bias magnetic fields.

Fig. 9. The frequency shift of the peak positions of mode 3 when a ferrite disk is loaded by dielectric samples.

Fig. 10. The frequency shift of the peak positions of mode 5 when a ferrite disk is loaded by dielectric samples.

Fig. 11. The frequency shift of the peak positions of mode 7 when a ferrite disk is loaded by dielectric samples.

Fig. 12. Dependences between $\varepsilon_r$ and a modulus of $\mu_a$ for modes 3, 5 and 7.

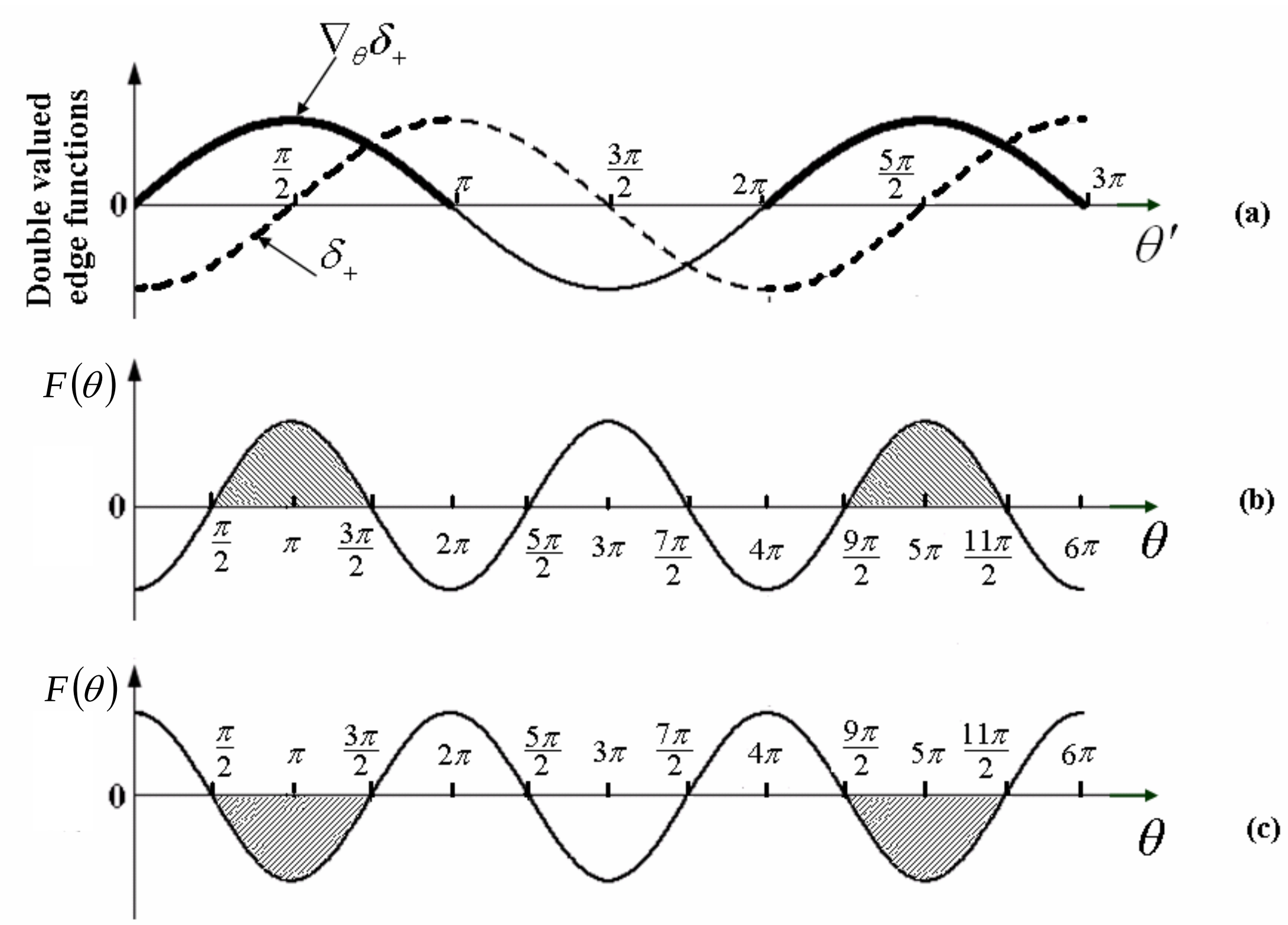

Fig. 1. Interaction of the double-valued-function edge state with the single-valued-function cavity electromagnetic fields in a case of the (+) resonance. (a) Double-valued edge function; (b) the cavity field at the "resonance absorption" interaction; (c) the cavity field at the "resonance repulsion" interaction.

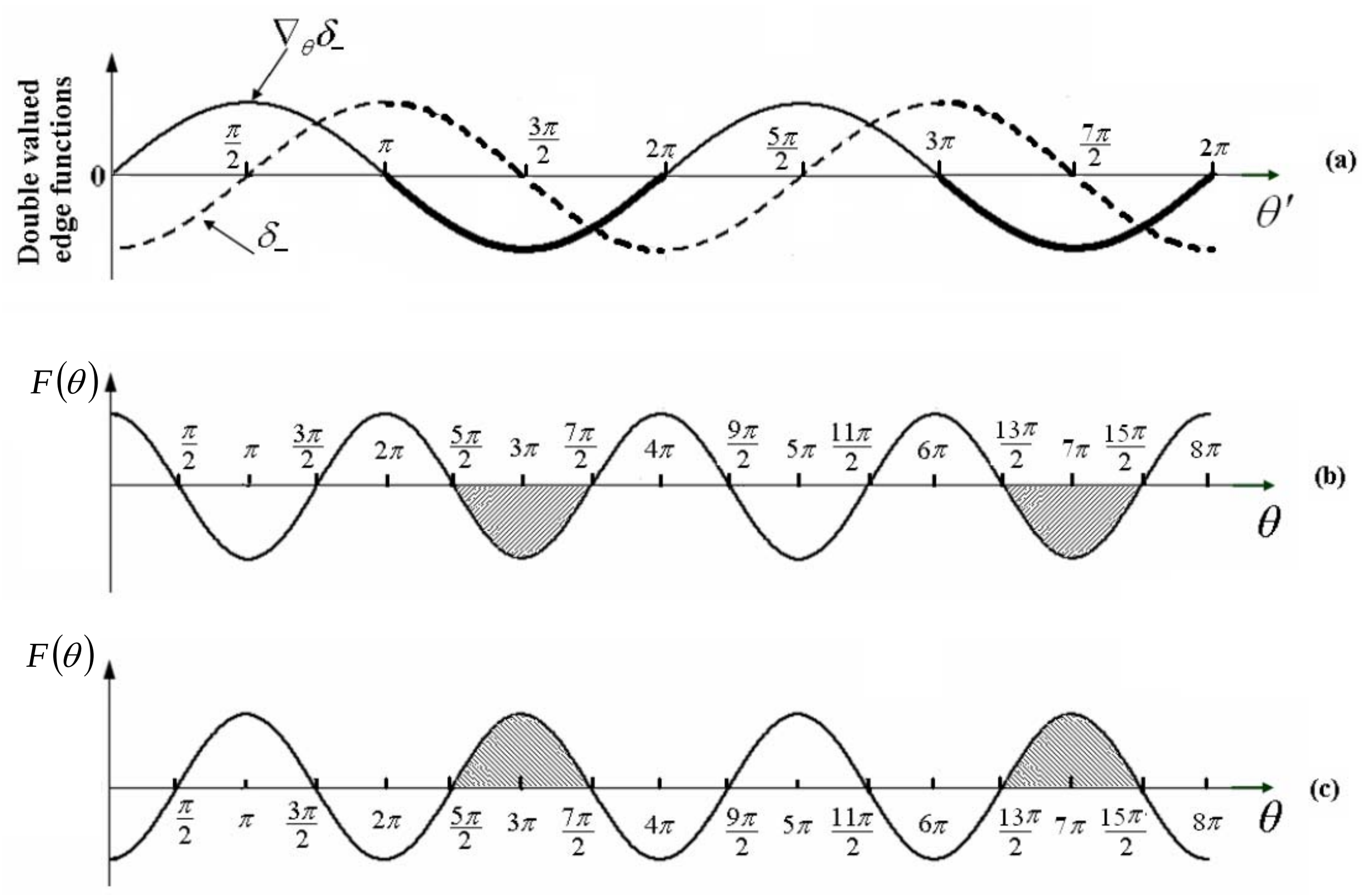

Fig. 2. Interaction of the double-valued-function edge state with the single-valued-function cavity electromagnetic fields in a case of the (–) resonance. (a) Double-valued edge function; (b) the cavity field at the "resonance absorption" interaction; (c) the cavity field at the "resonance repulsion" interaction.

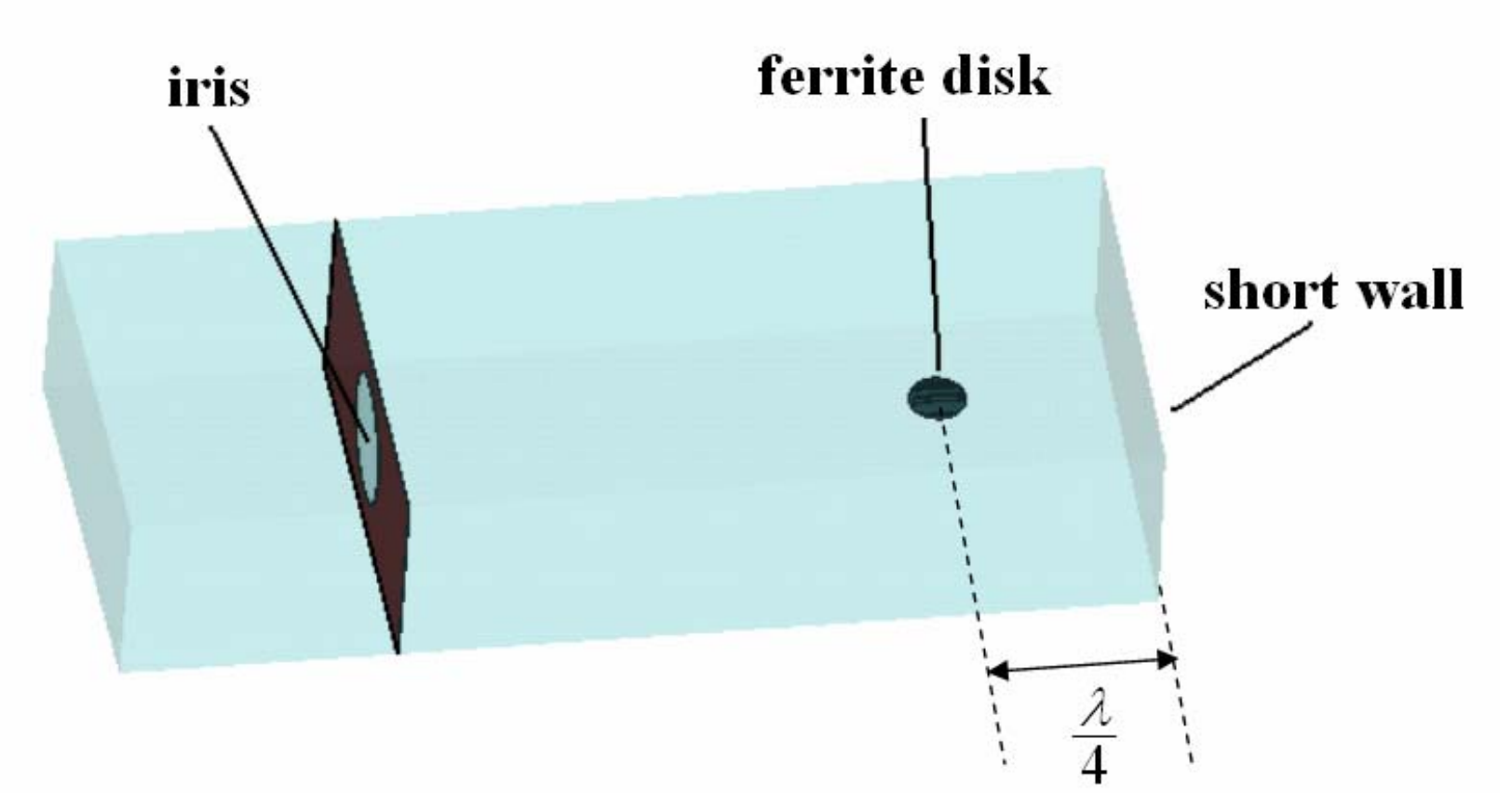

Fig. 3. Rectangular cavity with an inserted ferrite disk

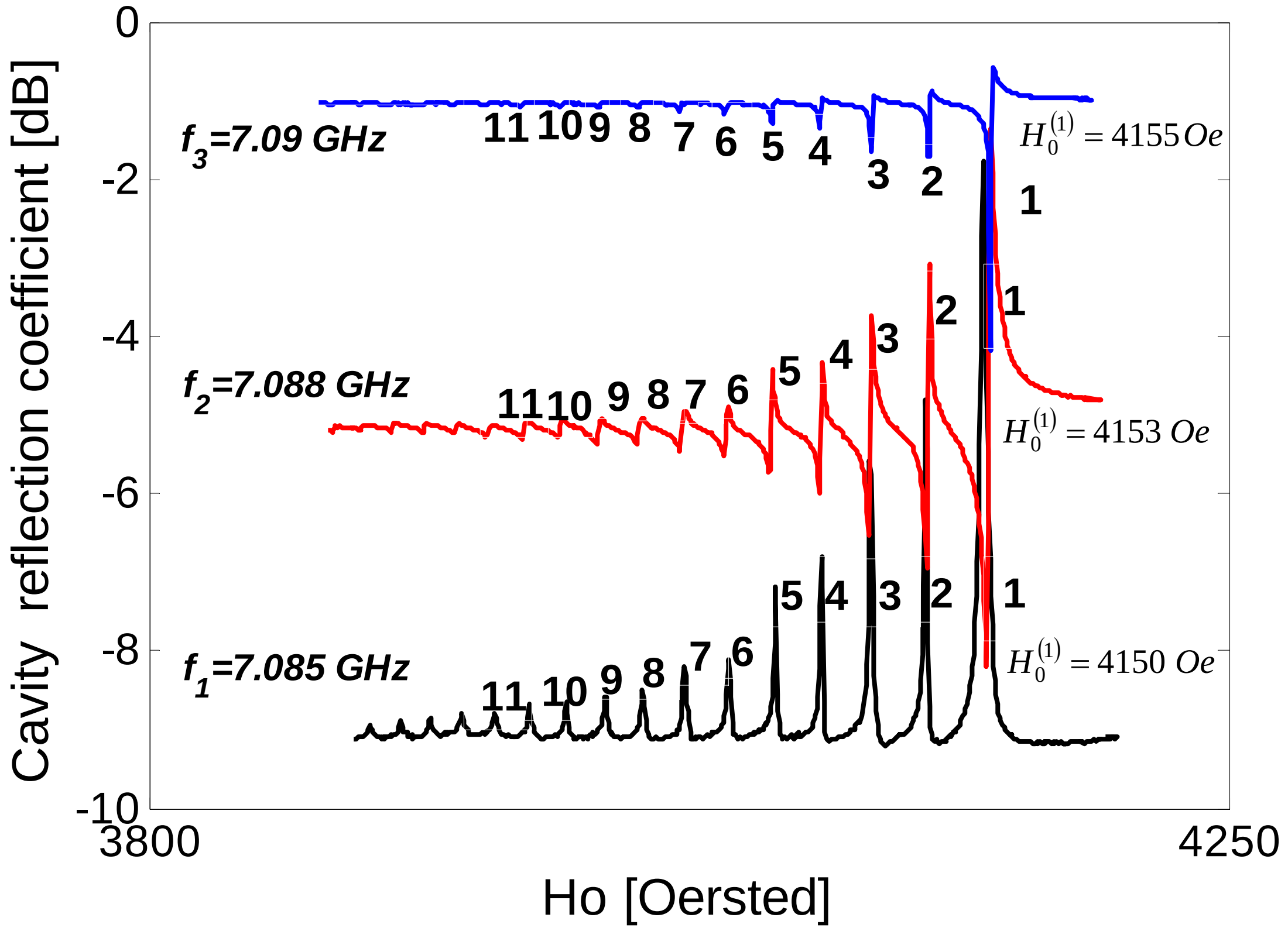

(*a*)

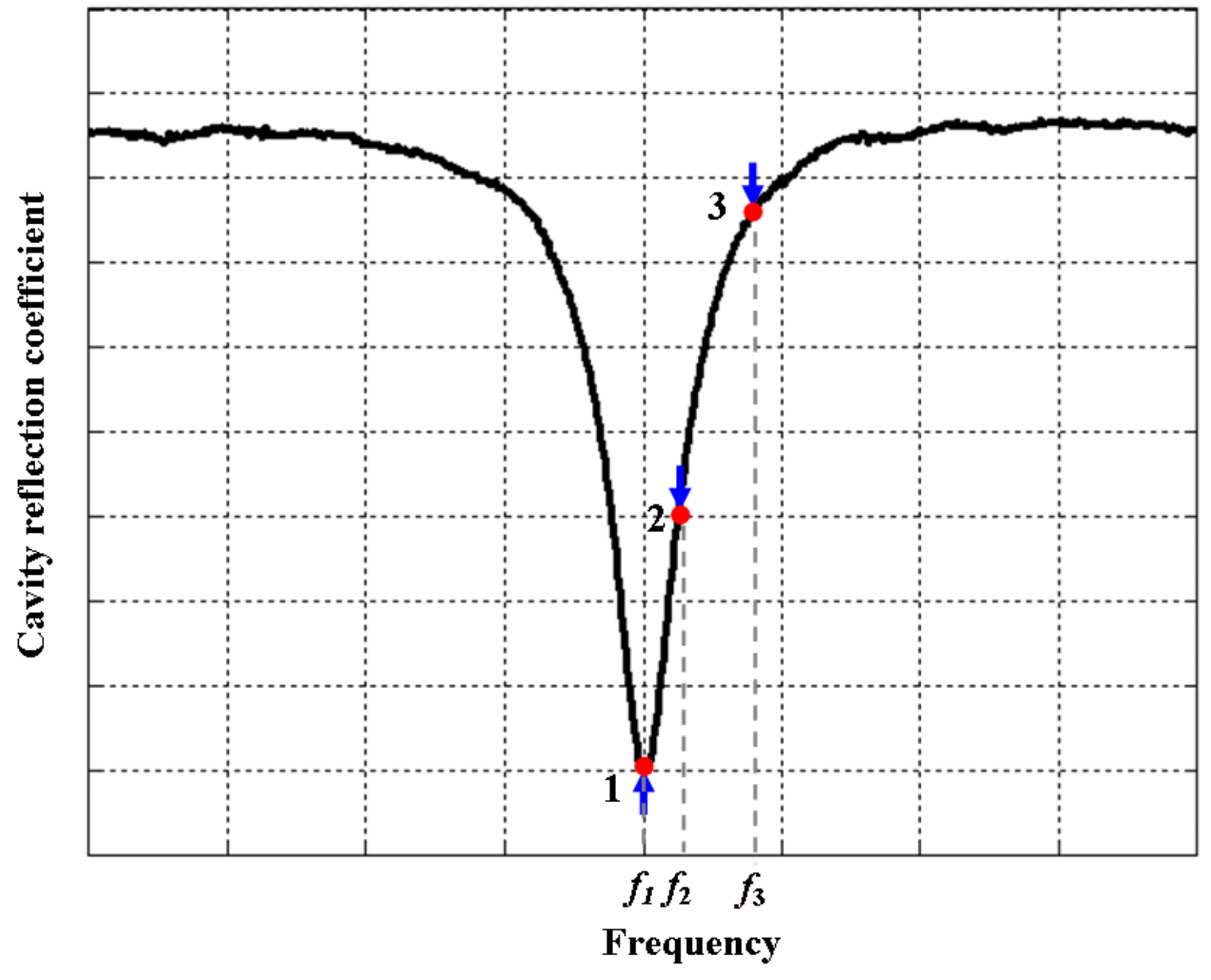

(*b*)

Fig. 4. Experimental evidence for the "resonance repulsion", "resonance absorption", and transitional behaviors. (a) The multiresonance spectral pictures for the "resonance repulsion" (at frequency $f_1$), "resonance absorption" (at frequency $f_3$), and transitional behavior (at frequency $f_2$); (b) frequencies $f_1$, $f_{2,}$ and $f_3$ on the resonance curve of the cavity.

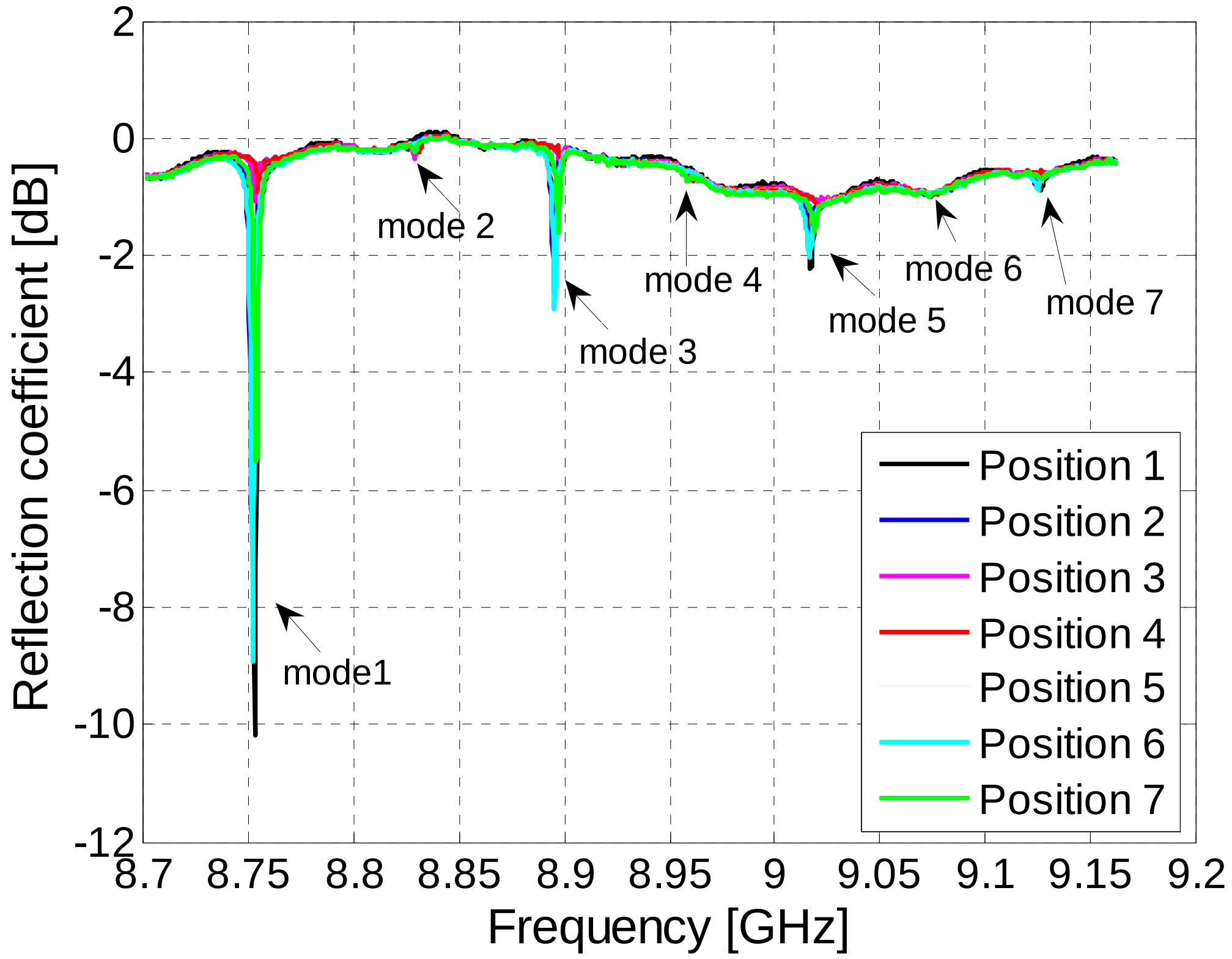

(*a*)

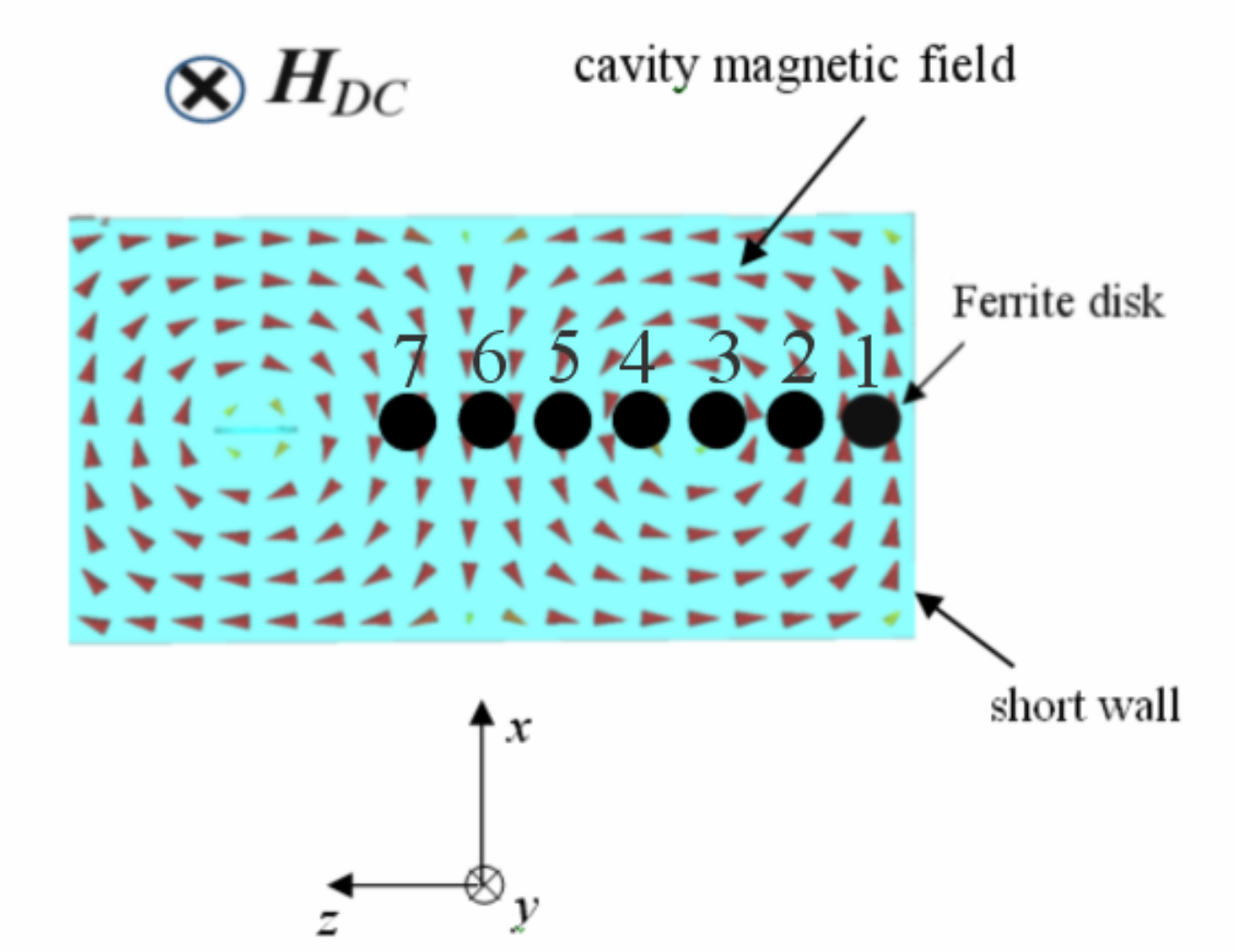

*(b)*

Fig. 5. The spectra of a ferrite disk without dielectric loading. (a) The MDM spectra for different disk positions; (b) disk positions in a waveguide.

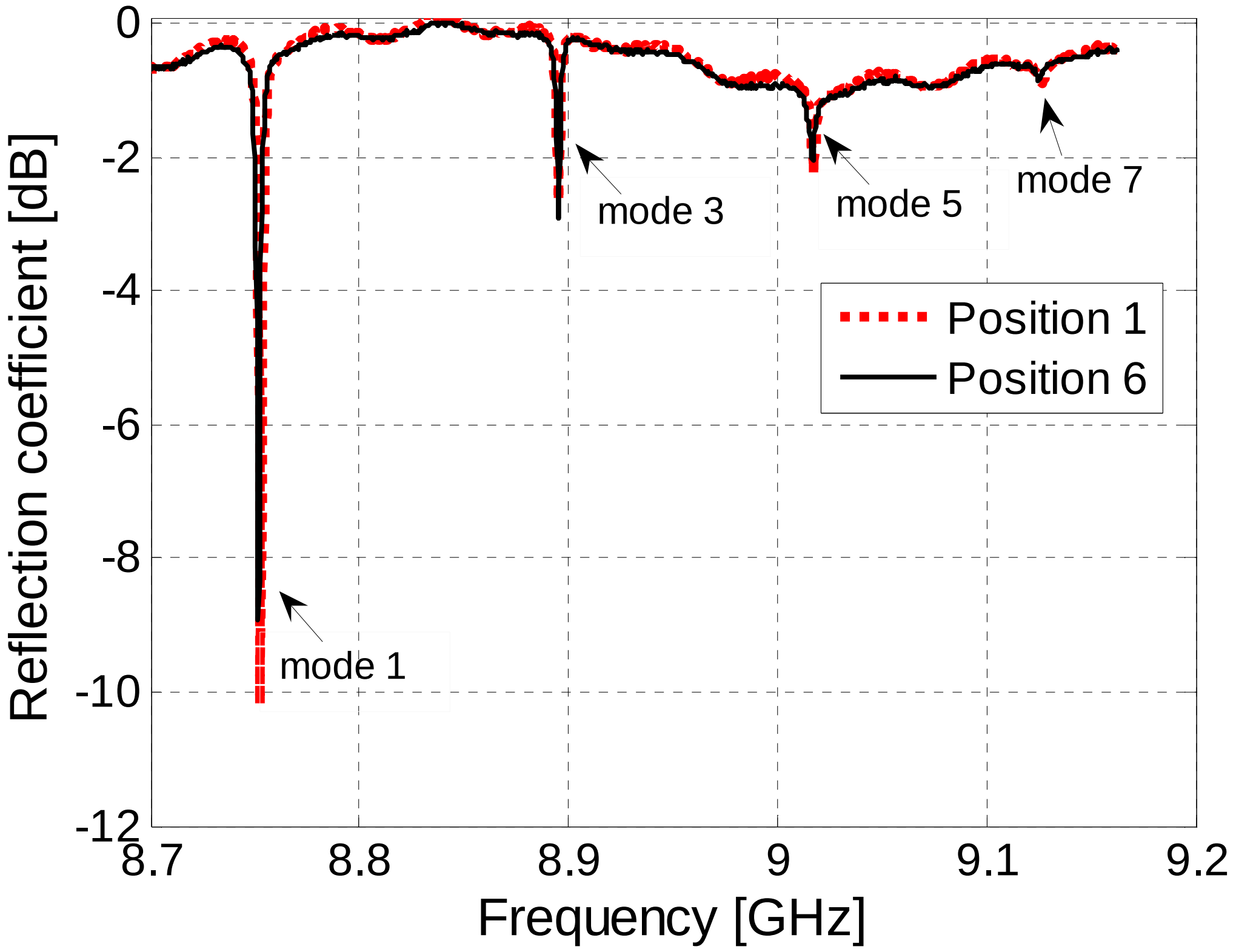

*(a)*

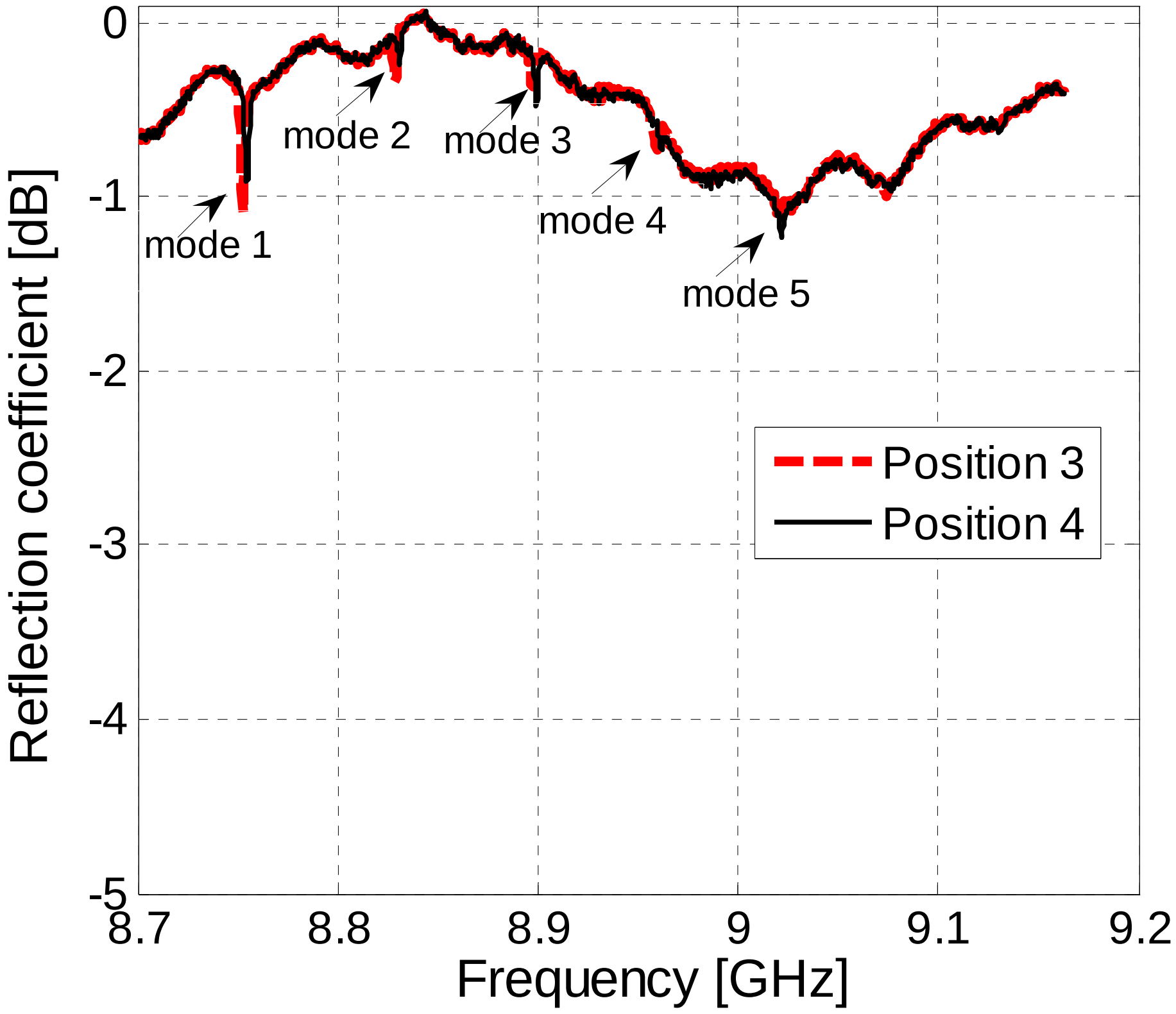

*(b)*

Fig. 6. The spectra of a ferrite disk without dielectric loading (detailed spectral pictures). (a) The MDM spectra for positions numbered 1 and 6; (b) the MDM spectra for positions numbered 3 and 4.

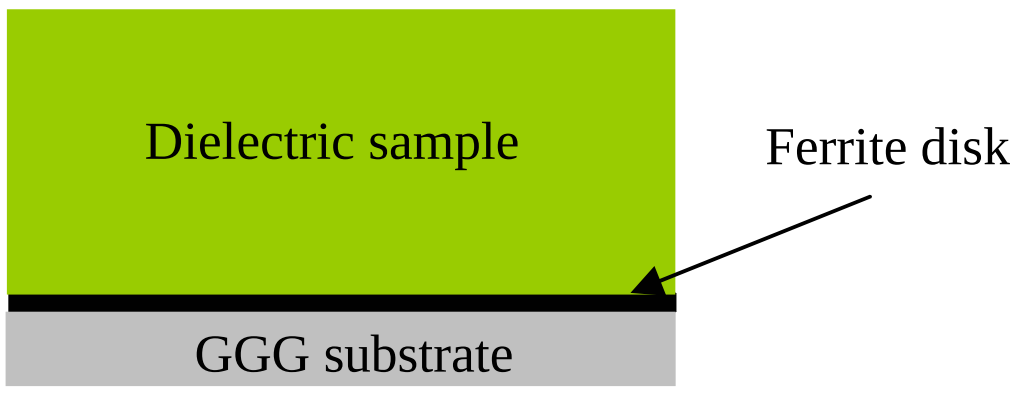

Fig. 7. A ferrite disk on a GGG substrate with a dielectric sample.

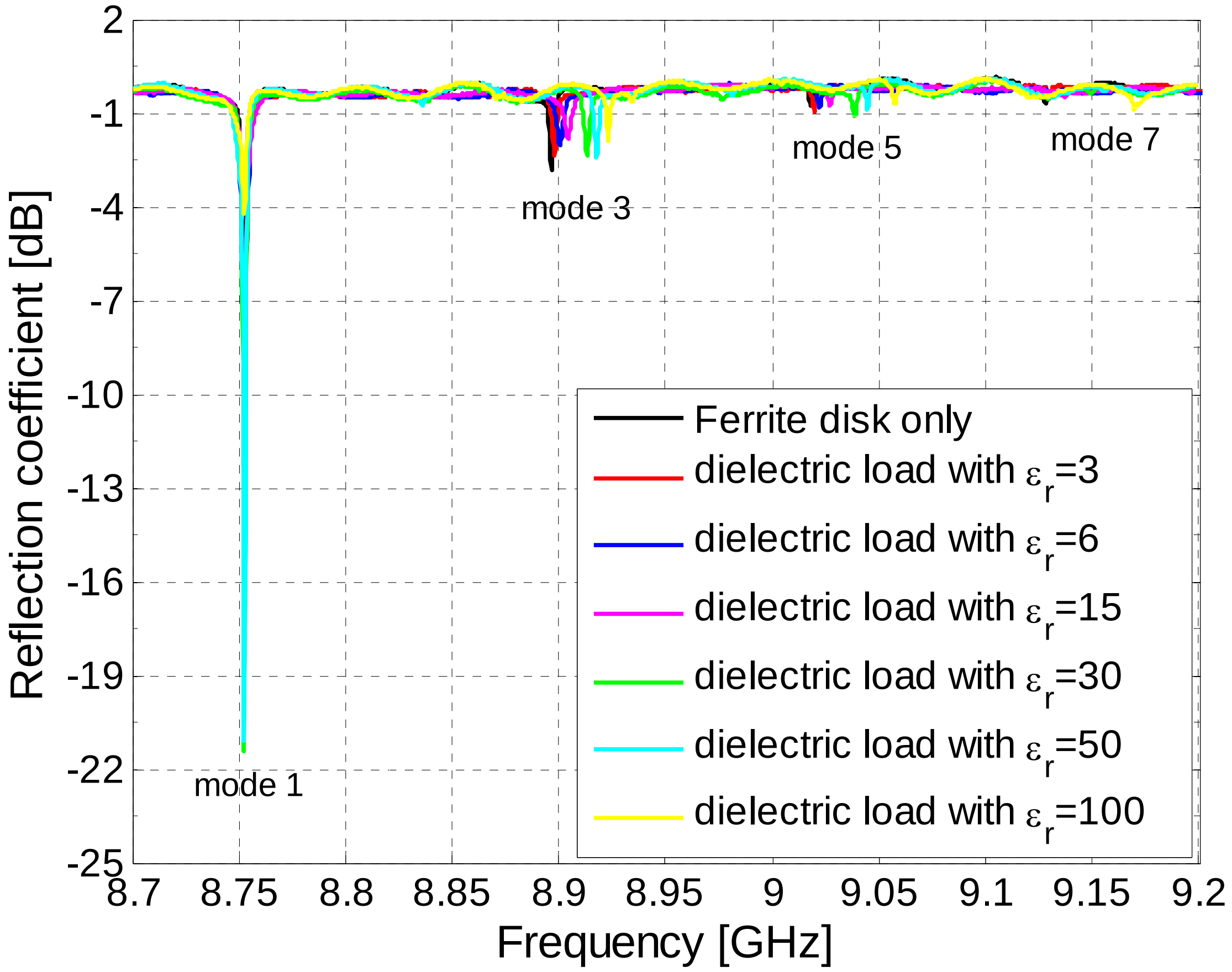

Fig. 8. The frequency shift of the mode peak positions for a ferrite disk with dielectric loading. The first peaks in the spectra are matched by proper correlations of bias magnetic fields.

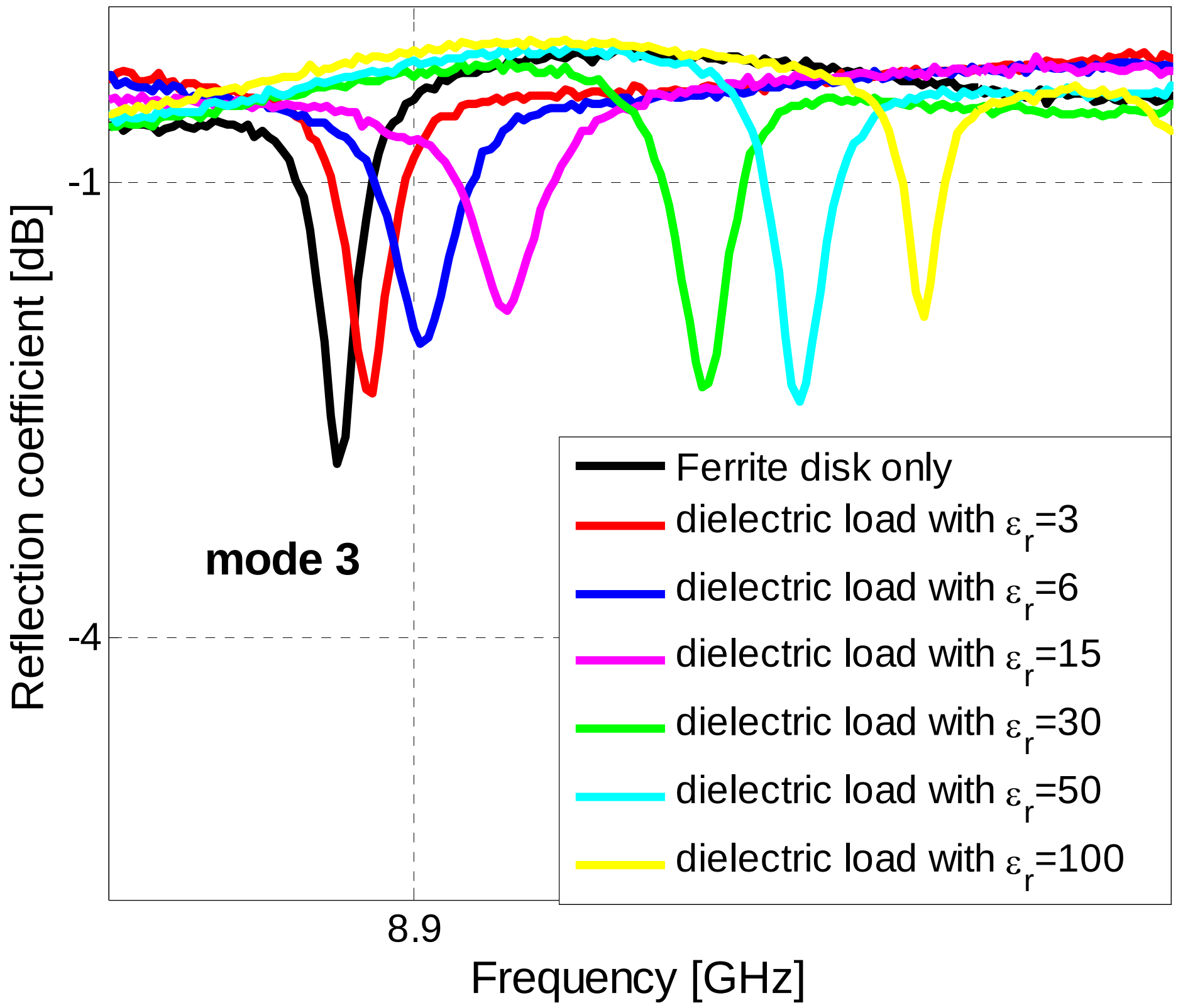

Fig. 9. The frequency shift of the peak positions of mode 3 when a ferrite disk is loaded by dielectric samples.

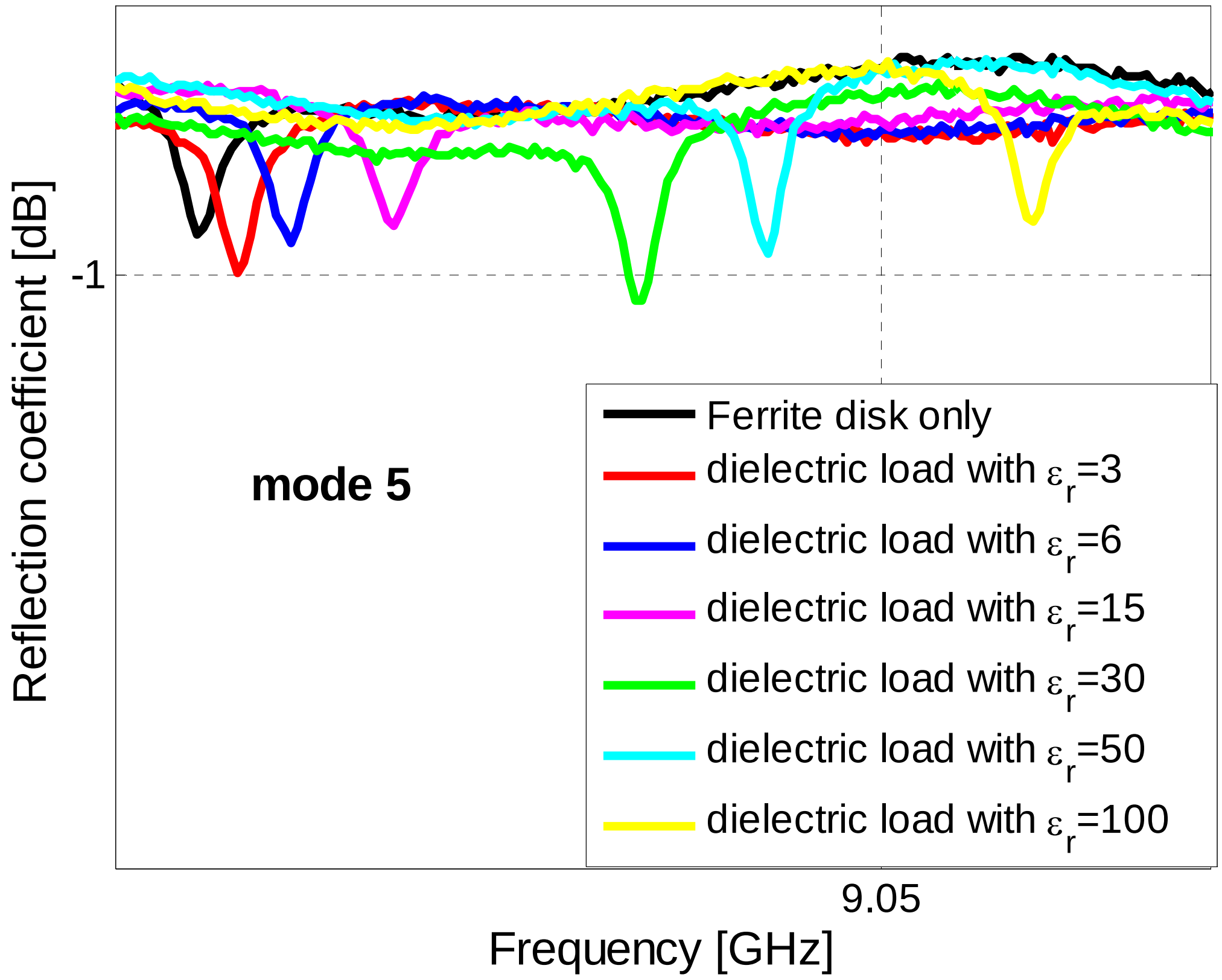

Fig. 10. The frequency shift of the peak positions of mode 5 when a ferrite disk is loaded by dielectric samples.

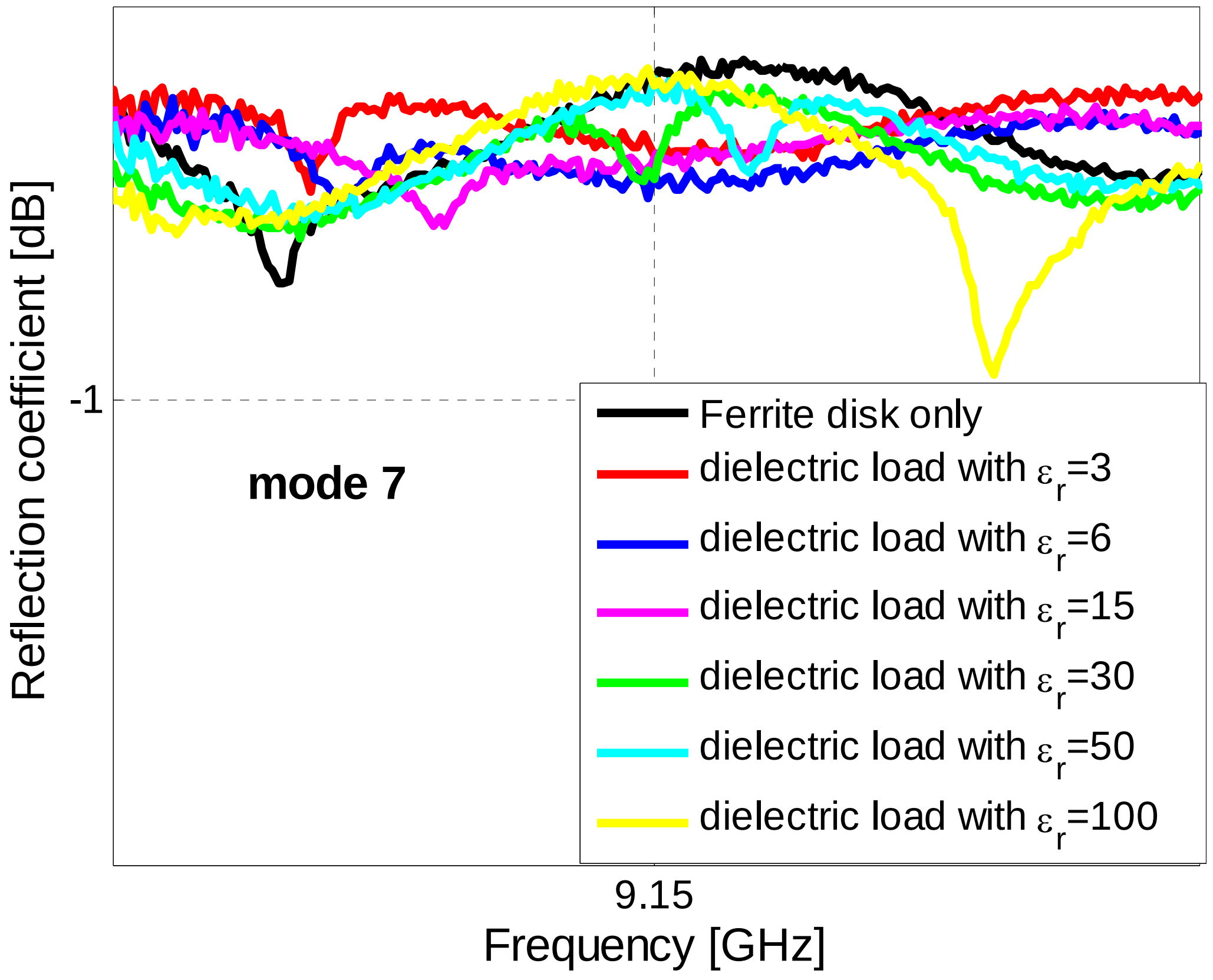

Fig. 11. The frequency shift of the peak positions of mode 7 when a ferrite disk is loaded by dielectric samples.

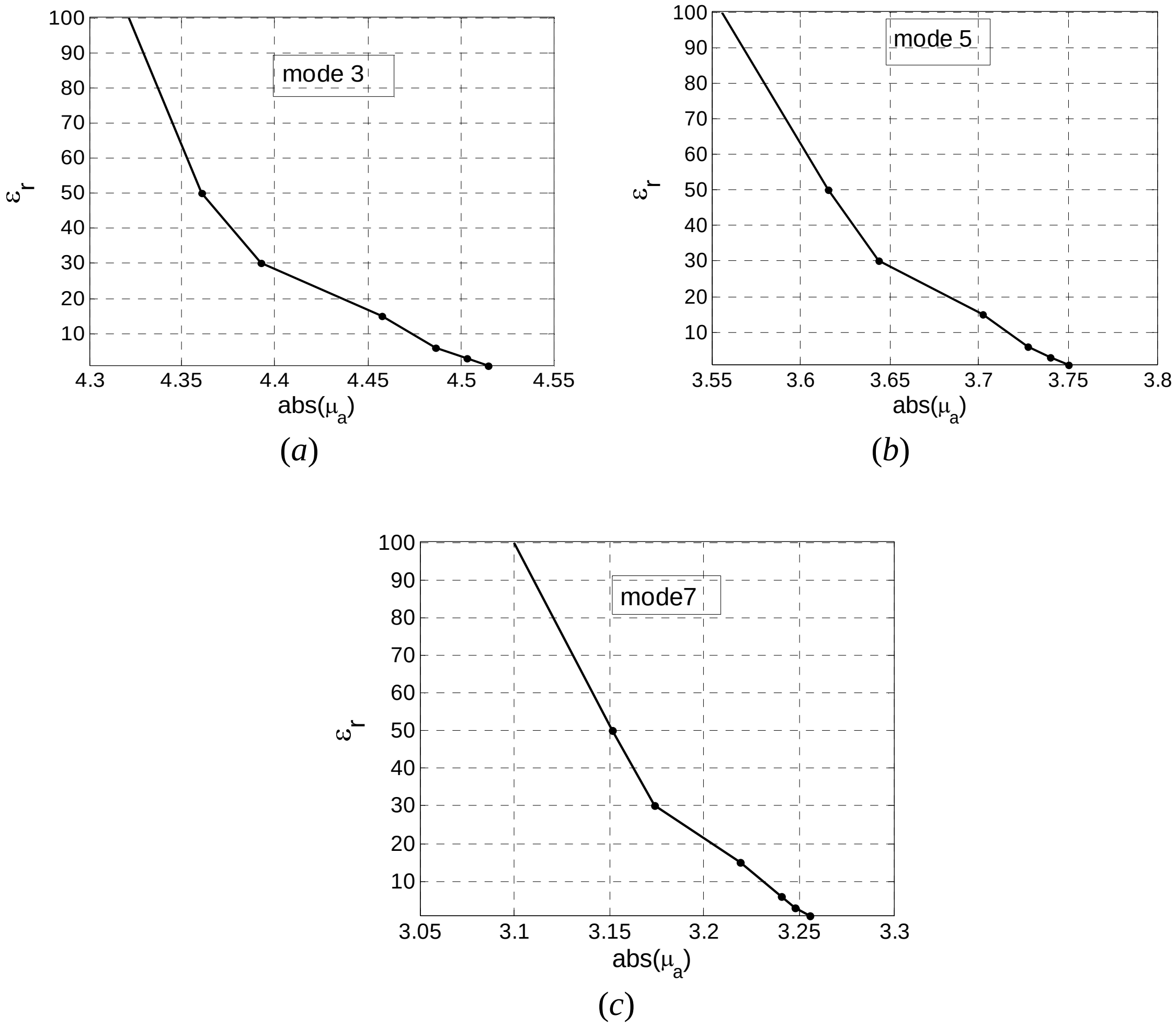

Fig. 12. Dependences between $\varepsilon_r$ and a modulus of $\mu_a$ for modes 3, 5 and 7.